\documentclass[prd,aps,preprint,showpacs,tightenlines]{revtex4}
\begin{document}
\preprint{}

\title{The Counting of Generalized Polarizabilities} 

\author{Richard F. Lebed}
\email{Richard.Lebed@asu.edu}
\affiliation{Department of Physics and Astronomy,
Arizona State University, Tempe, Arizona 85287 }

\date{May, 2001}

\begin{abstract}
We demonstrate a concise method to enumerate the number of generalized
polarizabilities---quantities characterizing the independent
observables in singly-virtual Compton scattering---for a target
particle of arbitrary spin $s$.  By using crossing symmetry and
$J^{PC}$ conservation, we show that this number is
$(10s+1+\delta_{s,0})$.
\end{abstract}


\maketitle
\thispagestyle{empty}

\setcounter{page}{1}

\section{Introduction}

The process of virtual Compton scattering (VCS) off a target $t$ of
spin $s$ offers new opportunities for probing the structure of $t$.
In particular, the singly-virtual process $\gamma^* + t \to \gamma +
t$ obtained through $e^- + t \to e^- + \gamma + t$ is represented by
more independent amplitudes than real Compton scattering (RCS) $\gamma
+ t \to \gamma + t$, owing to the additional longitudinal polarization
of the virtual photon, and contains a richer dynamics than RCS since
the virtual photon energy and momentum transfer are independent
variables.  Furthermore, the singly-virtual process is experimentally
feasible at laboratories such as Jefferson Lab, MIT-Bates, and MAMI.

The amplitudes for VCS are complicated to analyze because the real
photon may couple either to the electron (Bethe-Heitler processes) or
to the target $t$.  In addition, there are trivial Born diagrams in
which the intermediate state connecting the initial and final on-shell
target particles $t$ is just an off-shell copy of $t$; in such
diagrams the internal structure of $t$ enters only through its elastic
form factor at the vertices.

A theorem due to Low~\cite{Low} proves that all photon scattering
amplitudes from a target are regular once the Born terms have been
subtracted, and indeed are at least of linear order in the real photon
energy $\omega^\prime$.  The observable coefficients of these linear
terms characterize the scattering process in the limit of a soft
final-state photon, and are called {\it generalized polarizabilities}
(GPs).  While the separation of GPs from Bethe-Heitler and Born
amplitudes is a challenging problem of experimental analysis, the GPs
themselves contain interesting physical information.  The usual
polarizabilities measure the response of a target to external
electromagnetic fields, represented by scattering of photons with
$q^2=0$.  Generalized polarizabilities include $q$-dependence from the
photon virtuality and thus are momentum-space quantities with the same
relationship to the target polarization densities as the elastic
charge form factor has to the target charge density (i.e., a Fourier
transform into configuration space) \cite{Holstein}.

The question of counting the independent GPs of a target of a given
spin $s$ has been discussed in the literature for $s = 0$ or 1/2.  The
latter case of course includes nucleons, while the former describes
the spin-averaged case for any $s$.  Previously, the GPs were counted
by constructing~\cite{GLT} the Compton amplitudes associated with a
multipole expansion of the initial and final photons and matching
the total $J^P$ eigenvalues of the asymptotic states.  In particular,
the authors of Ref.~\cite{GLT} found 3 GPs for the spin-averaged case
and 10 for the full spin-1/2 case.

Subsequently, a direct calculation (Ref.~\cite{MD}) showed that only 2
of the 3 spin-averaged GPs are independent in the linear $\sigma$
model, raising suspicions that the GPs of Ref.~\cite{GLT} contain
redundancies.  In Ref.~\cite{DKMS} the source of the over-counting was
identified: The construction of Ref.~\cite{GLT} does not take into
account the crossing symmetry (combined with charge conjugation
symmetry) of the target $t$ in the process $\gamma^* + t \to \gamma +
t$, and direct construction of the relevant tensors in the Compton
scattering amplitude led to the expected 2 GPs in the spin-averaged
channel.  Subsequent work (Ref.~\cite{DKKMS}) by the same group showed
that only 6 GPs are independent in the $s=1/2$ case.

The latter counting poses several interesting questions.  First, is it
obvious that the new construction does indeed take into account all
the relevant symmetries?  Second, this construction, while elegant, is
still rather formidable and tends to obscure a direct understanding of
why the counting of GPs turns out to give a particular number.  Third,
there is a very interesting numerical relation that holds for $s=0$
and 1/2: The number of GPs equals in each case (2 and 6, respectively)
the number of amplitudes for RCS!  We hasten to add that, even {\it a
priori}, this is almost certainly a coincidence, since the GPs are
associated with amplitudes linear in $\omega^\prime$ for an arbitrary
$q^2$ in a fixed partial wave, while the real Compton amplitudes have
$q^2=0$ and an arbitrary $\omega^\prime$, but sum over all allowed
partial waves.  If a one-to-one correspondence existed between these
two disparate quantities, it would necessarily point to a highly
subtle feature of the full Compton $S$-matrix.

In this short paper we seek to impose all relevant symmetries to the
express end of enumerating the GPs for a target of arbitrary spin $s$.
As in Refs.~\cite{DKMS,DKKMS} we find that crossing symmetry and
charge conjugation symmetry are indeed essential ingredients to obtain
the correct counting, and we render these symmetries manifest by
performing the counting for the crossed process $\gamma^* + \gamma \to
t + \bar t$.  We obtain the same counting as these authors for $s=0$
and 1/2.  We also compute, for comparison, the number of amplitudes
obtained for $\gamma^* + t \to \gamma + t$ through direct counting of
partial waves in the multipole expansion (as in Ref.~\cite{GLT}), or
equivalently, through helicity amplitudes: It is interesting to verify
and extend the result of Ref.~\cite{GLT}, and count the number of
potential GPs eliminated by crossing symmetry.  Finally, we perform a
counting of helicity amplitudes in RCS---hardly a new problem, but
included for completeness---and show that this number is larger than
the number of GPs for targets with $s \ge 1$.

Apart from its intrinsic interest as a problem of separating purely
symmetry-related and dynamical degrees of freedom, this counting
proves useful if one wishes to measure the GPs of composite targets.
For example, given the complete set of GPs of the deuteron (we show
that there are 11) and the 6 of the proton, it is possible to obtain
interesting information on the electromagnetic structure of the
neutron.

\section{GP Counting Without Crossing Constraints}
\subsection{The Multipole Basis}

We begin by noting the central constraint that makes the counting of
GPs different from that of full helicity amplitudes, namely, that the
final-state photon is assumed soft.  In particular, this implies that
only the E1($1^-$) and M1($1^+$) multipoles are permitted, since they
alone have wave functions for transversely polarized photons that vary
linearly with the photon energy $\omega^\prime$; the higher multipoles
are at least quadratic in $\omega^\prime$.  Thus, only a limited set
of relative angular momenta between the final-state $t$ and $\gamma$
are allowed, which are already included in the multipole wave
function.

To be explicit, let us work in the center of momentum (CM) frame of
the process $\gamma^*(q) + t(p) \to \gamma(q^\prime) + t(p^\prime)$,
which we call the $s$ channel.  Without loss of generality, the
momentum components in this frame (using the notation of
Refs.~\cite{DKMS,DKKMS}) may be labeled
\begin{eqnarray}
q^\mu & = & (\omega_0, 0, 0, +\bar q) \ , \nonumber \\
p^\mu & = & (E       , 0, 0, -\bar q) \ , \nonumber \\
q^{\prime \, \mu} & = & (\omega^\prime  ,  +\omega^\prime \sin \theta,
0, +\omega^\prime \cos \theta) \ , \nonumber \\
p^{\prime \, \mu} & = & (E^\prime       ,  -\omega^\prime \sin \theta,
0, -\omega^\prime \cos \theta) \ ,
\end{eqnarray}
where $E = \sqrt{\bar q^2 + m_t^2}$, $E^\prime = \sqrt{\omega^{\prime
\, 2} + m_t^2}$, and overall energy conservation requires $\omega_0 =
E^\prime + \omega^\prime - E$.  Let us also define $P = p + p^\prime$.
There are clearly three independent kinematic variables, whether
labeled as $\omega^\prime$, $\bar q$, and $\theta$, or the invariants
$q^2$, $q \cdot q^\prime$, and $q \cdot P$, or the Mandelstam
variables $s$, $t$, and $u$ (note that $s+t+u$ is not fixed since the
virtual photon does not have a fixed mass).  In particular, one
finds
\begin{equation} \label{omegas}
\omega^\prime = \frac{q \cdot q^\prime + q \cdot P}{2 \sqrt{m_t^2 + q
\cdot q^\prime + q \cdot P}} = \frac{s-m_t^2}{2\sqrt{s}} \ .
\end{equation}

The counting of allowed partial wave amplitudes in the $s$ channel may
be carried out either using the multipole expansion or helicity
amplitudes, and we now demonstrate how each is accomplished.

To find the total $J^P$ of a given final state, one need only add the
E1 or M1 photon to the target spin-parity, $s^\Pi$.  These values are
$J^P = \{|s-1|, \ldots, s+1\}^{-\Pi}$ and $\{|s-1|, \ldots,
s+1\}^\Pi$, respectively.  These sets contain 3 values of $J$ if $s
\ge 1$, 2 if $s=1/2$, and 1 if $s=0$, and so it is convenient to treat
the exceptional cases $s=0$ and 1/2 explicitly.  The initial virtual
photon may be in any multipole (including the longitudinal Coulomb
modes) for which adding the initial target spin-parity $s^\Pi$ gives
the same total $J^P$.  For example, if $s^\Pi = 1^+$ and the final
photon is E1, then $J^P = 0^-, 1^-, 2^-$.  Among the initial allowed
multipoles is M2($2^-$), which added to $1^+$ gives $1^-, 2^-, 3^-$.
Thus one finds two allowed amplitudes for an initial M2 photon and
final E1 photon, corresponding to $J^P = 1^-, 2^-$.  By the triangle
rule of angular momentum addition, the largest allowed multipole rank
$n$ satisfies $n-s = s+1$, or $n=2s+1$.

Since allowing both E1 and M1 photons includes both parities, generic
values of $J$ permit an equal number of amplitudes coupled to E$n$,
M$n$, and C$n$.  The exception is for $n=0$, since there are no E0 or
M0 multipoles but only C0($0^+$).  In that case, the only potentially
allowed amplitude has $J^P = s^\Pi$, which is always permitted by the
final state unless $s = 0$.  Thus there are $(1-\delta_{s,0})$
amplitudes with initial C0 photons.  All other values of $n$ give 3
amplitudes.  The total number is therefore
\begin{equation}
(1-\delta_{s,0}) + 3 \sum_{n=1}^{2s+1} \left[ \mbox{Overlaps between}
\ \{ |n-s|, \ldots, n+s \} \ \mbox{and} \ \{ |s-1|, \ldots, s+1 \}
\right] \ .
\end{equation}
It is straightforward to count the terms in the sum by dividing them
into cases.  The members of the set $\{ |s-1|, \ldots, s+1 \}$ are
distinct if $s \ge 1$, leading to 3 overlaps for $n \le 2s-1$, 2 for
$n=2s$, and 1 for $n=2s+1$; the special cases $s=0$ and 1/2 may be
handled separately. Carrying this out, one finds the number of
amplitudes in the $s$ channel to be
\begin{equation} \label{GPold}
18s + 1 + 2\delta_{s,0} \ .
\end{equation}
We hasten to add that this is the counting of amplitudes not taking
into account the crossing symmetry of the initial and final $t$
particles.  Only $J^P$ conservation and the exhaustion of
$O(\omega^\prime)$ amplitudes by E1 and M1 multipoles has been used.
This is, in fact, a counting equivalent to that done in
Ref.~\cite{GLT}, that of counting the number of coefficients of terms
linear in $\omega^\prime$, identified in Ref.~\cite{GLT} as the GPs:
The cases $s=0$ and 1/2 give 3 and 10, respectively.

\subsection{The Helicity Amplitude Basis}

One might worry that the multipole expansion is relevant only
nonrelativistically (since $\omega^\prime$ is treated as a
perturbative parameter).  However, the multipoles simply count quantum
numbers and therefore the corresponding amplitudes are fully
relativistic.  Put another way, the true multipole amplitudes begin
with a certain power of $\omega^\prime$ as determined by their
low-energy behavior, but contain higher powers as well.  One can
verify this statement explicitly by employing a manifestly
relativistic formalism, the helicity decomposition~\cite{JW,landau}.
The counting procedure is very similar to that performed in
Ref.~\cite{JL}.  Define $\psi_{JM\lambda_1 \lambda_2}$ as a
two-particle helicity state in the CM.  Angular momentum conservation
along the CM axis requires $|\lambda_1 - \lambda_2| \le J$.  With
$\eta_{1,2}$ being the intrinsic parities of the two particles, the
action of the parity operator is
\begin{equation} \label{P}
\hat P \psi_{JM\lambda_1 \lambda_2} = \eta_1 \eta_2 (-1)^J \psi_{JM
-\lambda_1 -\lambda_2} \ ,
\end{equation}
meaning that the parity $\pm$ eigenstates are given by
\begin{equation} \label{parity}
\psi_{JM\lambda_1 \lambda_2} \pm \eta_1 \eta_2 (-1)^J \psi_{JM
-\lambda_1 -\lambda_2} \ .
\end{equation}

It is not necessary to carry out a helicity decomposition of the final
state since the restriction to the E1 and M1 multipoles already
determines the allowed values of $J^P$; indeed, the assumption of a
soft photon already lies beyond the scope of the helicity
decomposition, which is completely general since it allows {\em any}
relative angular momentum.  On the other hand, one may study the
initial state of $\gamma^* + t$ in terms of helicity basis partial
waves instead of using the multipole basis as in the previous
counting.  The intrinsic parity of the virtual photon is $-1$; only
the $1^-$ polarizations need be considered, since the $0^+$
polarization may be eliminated through current conservation.  Thus,
the product $\eta_1 \eta_2$ in Eq.~(\ref{parity}) equals $-\Pi$.
Moreover, the two terms in Eq.~(\ref{parity}) are distinct, so that
both parity eigenstates survive, unless $\lambda_1 = \lambda_2 = 0$,
in which case only one parity eigenstate survives for each value of
$J$.

Now the physical input is complete, and one need only enumerate
possible values of $\lambda$ to count amplitudes.  Since each
combination in Eq.~(\ref{parity}) includes both the $\lambda_1 ,
\lambda_2$ and $-\lambda_1 , -\lambda_2$ combinations, it is
completely general to consider only the two types of combinations ($0
< \lambda_1 \le s_1$, $-s_2 \le \lambda_2 \le s_2$) or $(\lambda_1 =
0$ [if 1 is a boson], $0 \le \lambda_2 \le s_2$).  In the present
case, let 1 be the virtual photon so that $s_1=1$, and 2 be the target
so that $s_2=s$.  Then the two cases are ($\lambda_{\gamma^*} = +1$,
$-s \le \lambda_t \le s$) and ($\lambda_{\gamma^*} = 0$, $\lambda_s
\ge 0$).

If $\lambda_{\gamma^*} = +1$, then the restriction on $J$ reads
$|1-\lambda_t| \le J \in \{ |s-1|, s, s+1 \}$.  We have seen that the
parity eigenvalue does not matter here; if $\lambda_t$ is allowed then
both parities appear.  The special cases $s=0$ and 1/2 may again be
handled separately, so that the members of $\{ s-1, s, s+1 \}$ are
distinct.  Then, by dividing into classes based on whether
$|1-\lambda_t| \le s-1$ (leading to 3 amplitudes), or
$|1-\lambda_t|=s$ (2 amplitudes), or $|1-\lambda_t|=s+1$ (1
amplitude), one finds a total of $12s$ amplitudes, including parity
doubling.

For $\lambda_{\gamma^*} = 0$, only $\lambda_t \ge 0$ need be
considered, and only $\lambda_t \neq 0$ receives a parity doubling.
Again handling $s=0$ and 1/2 separately, one may take $s \ge 1$.  Then
the restriction on $J$ reads $\lambda_t \le J \in \{ s-1, s, s+1 \}$,
leading to 3 amplitudes ($\lambda_t \le s-1$) or 2 amplitudes
($\lambda_t = s$).  With the floor function denoted as usual by square
brackets, there are $[s+1/2]$ values of $\lambda_t > 0$ and
$[1+(-1)^{2s}]/2$ values of $\lambda_t = 0$, and weighting these by
the parity multiplicities, one finds a total of $6s+1$ amplitudes [all
factors of $(-1)^{2s}$ cancel].  Combining this with the
$\lambda_{\gamma^*} = +1$ amplitudes and including the special cases
of $s=0$ and 1/2 (which give the same number of amplitudes as before),
one again finds a total of $(18s+1+2\delta_{s,0})$ amplitudes.

\section{GP Counting Using Crossing Constraints}

As mentioned above, the $s$ channel is not the channel of highest
symmetry because there are constraints due to the crossing symmetry
that have not been taken into account.  We accomplish this by
considering instead the $t$ channel process $\gamma^* (q) + \gamma
(-q^\prime) \to t(p^\prime) + \bar t (-p)$ (the double use of $t$ as a
kinematic variable and the target particle should not confuse the
reader).  The signs on the momenta are designed to allow the same
notation for all dot products in the $S$ matrix regardless of which
crossed channel one considers.  That is, $p+q = p^\prime + q^\prime$
in all channels.

It is again convenient in the helicity decomposition to consider
the process in the CM.  The momentum components may be labeled
\begin{eqnarray}
q^\mu              & = & (q_0   , 0, 0, +\omega) \ , \nonumber \\
-q^{\prime \, \mu} & = & (\omega, 0, 0, -\omega) \ , \nonumber \\
p^{\prime \, \mu}  & = & (\bar E^\prime, -\rho \sin \phi, 0,
-\rho \cos \phi) \ , \nonumber \\
-p^\mu             & = & (\bar E, +\rho \sin \phi, 0, +\rho \cos \phi)
\ ,
\end{eqnarray}
where $\bar E = \sqrt{\rho^2 + m_t^2} = \bar E^\prime$ and energy
conservation gives $q_0 + \omega = 2\bar E$.  The three independent
kinematic variables in this frame may be taken as $\omega$, $\rho$,
and $\phi$, and one finds
\begin{equation} \label{omegat}
\omega = \frac{-q \cdot q^\prime}{\sqrt{q^2 - 2 q \cdot q^\prime}} =
\frac{2m_t^2-s-u}{2\sqrt{t}} \ .
\end{equation}

In both the $s$ and $t$ channels the real photon is assumed soft.
However, this implies in the $s$ channel CM that the final $t$ has a
small spatial momentum and in the $t$ channel that the $\gamma^*$ has
a small spatial momentum.  The two frames have different kinematics
[as is apparent from comparing Eqs.~(\ref{omegas}) and
(\ref{omegat})], so how can it be that there is a relationship between
amplitudes linear in $\omega^\prime$ and those linear in $\omega$?
The answer is that the transformation between $\omega^\prime$ and
$\omega$, whether written in terms of the other two kinematic
variables of the $s$ channel or the $t$ channel, preserves the
linearity in either variable.  Explicitly,
\begin{equation} \label{transf}
\omega = -\omega^\prime \frac{(E^\prime - E + \omega^\prime - \bar q
\cos \theta)} {\sqrt{2(m_t^2 - E^\prime E + \bar q \omega^\prime \cos
\theta)}} \ .
\end{equation}
This linear-to-linear mapping arises from the masslessness of the real
photon, and in particular that all components of $q^{\prime}$ vanish
like $\omega^\prime$ or $\omega$ in the soft photon limit.  It implies
that amplitudes linear in soft photon energies may be regarded as GPs
in either crossing, and that the process with the smaller number of
such amplitudes (call this the {\it minimal process}) provides the
correct counting of independent GPs.  The extra amplitudes in other
crossings must therefore be redundant precisely because of the ignored
constraints of crossing symmetry and the application of additional
symmetries manifest only in the minimal process.

Two points should be clarified before proceeding.  First, the
parameters $\omega$ or $\omega^\prime$, while each defined in the
relevant CM frame, actually represent two different functions of
Lorentz invariant quantities [as apparent from Eqs.~(\ref{omegas}) and
(\ref{omegat})].  They both become small in the soft photon limit
because two independent Lorentz invariants, $q \cdot q^\prime$ and $q
\cdot P$, or equivalently $s-m_t^2$ and $u-m_t^2$, become small in
this limit; the coefficients of both contribute to the GPs.  Second,
we note that the transformation Eq.~(\ref{transf}) is pure imaginary
since the argument of the denominator square root is negative for real
allowed values of $\bar q$, $\omega^\prime$, and $\theta$; this simply
indicates that fixed values for the three invariants that give real
kinematics in the $s$ channel do not give real kinematics in the $t$
channel.  But the existence of a certain number of amplitudes or GPs
should not depend on the reality of the momentum components, since the
full amplitudes are complex analytic functions of the invariants.

We now argue that the $t$ channel represents the minimal process.  We
have already noted that the $s$ channel possesses a crossing symmetry
between the initial and final $t$ not yet properly taken into account.
In the $t$ channel this symmetry is translated into charge conjugation
symmetry of the $t + \bar t$ system; the $\gamma^* + \gamma$ system
always has $C = (-1)^2 = +1$.  $C$ invariance is of course not
applicable in the $s$ channel unless $t$ is self-conjugate.  There is
clearly no remaining crossing symmetry to take into account in the $t$
channel, nor $T$ invariance, since the initial and final states
consist of completely distinct particles.  Thus, the only symmetry
restrictions applicable to the $t$ channel are $J^{PC}$ conservation.

As before, we work in the helicity formalism.  The action of charge
conjugation on a self-conjugate two-particle state gives
\begin{equation} \label{C}
\hat C \psi_{JM\lambda_1 \lambda_2} = (-1)^J \psi_{JM \lambda_2
\lambda_1} \ .
\end{equation}
Combining Eqs.~(\ref{P}) and (\ref{C}) and suppressing the $JM$
subscripts, one obtains the $J^{PC}$ eigenstates
\begin{eqnarray}
PC=++:~~~ & & \psi_{\lambda_1 \lambda_2} + (-1)^J \psi_{\lambda_2
\lambda_1} + (-1)^J \psi_{-\lambda_1 -\lambda_2} + \psi_{-\lambda_2
-\lambda_1} \ , \nonumber \\
PC=+-:~~~ & & \psi_{\lambda_1 \lambda_2} - (-1)^J \psi_{\lambda_2
\lambda_1} + (-1)^J \psi_{-\lambda_1 -\lambda_2} - \psi_{-\lambda_2
-\lambda_1} \ , \nonumber \\
PC=-+:~~~ & & \psi_{\lambda_1 \lambda_2} + (-1)^J \psi_{\lambda_2
\lambda_1} - (-1)^J \psi_{-\lambda_1 -\lambda_2} - \psi_{-\lambda_2
-\lambda_1} \ , \nonumber \\
PC=--:~~~ & & \psi_{\lambda_1 \lambda_2} - (-1)^J \psi_{\lambda_2
\lambda_1} - (-1)^J \psi_{-\lambda_1 -\lambda_2} + \psi_{-\lambda_2
-\lambda_1} \ . \label{hel}
\end{eqnarray}

As before, the soft photon appears only in the multipoles E1 and M1;
however, note that the relative angular momentum in this case is
measured with respect to the $\gamma^*$ rather than the $t$.  The
allowed $J^{PC}$ values are then $(0,1,2)^{++}$ for E1 and
$(0,1,2)^{-+}$ for M1.  For the $t + \bar t$ combination, therefore,
one need consider only the $J^{-+}$ and $J^{++}$ combinations.

The physics input now being complete, let us review the ingredients
before proceeding with the counting.  As before, start with $J^P$
conservation and the observation that only E1 and M1 multipoles give
CM amplitudes linear in $\omega^\prime$, and hence GPs.  Then note
that the original ($s$-channel) process possesses a crossing symmetry
due to the on-shell $t$ in both initial and final states.  Considering
instead the corresponding $t$-channel process, one finds that the
original crossing symmetry is no longer manifest; but unlike in the
$s$ channel, one may classify amplitudes by eigenvalues of charge
conjugation: Effectively, crossing symmetry has been traded for $C$
symmetry.  To complete the argument, it must be checked that GPs in
the $s$ channel (coefficients of amplitudes linear in $\omega^\prime$)
map to what may be called the GPs in the $t$ channel CM (coefficients
of amplitudes linear in $\omega$); this is verified explicitly by
Eq.~(\ref{transf}).  Since $J^{PC}$ conservation accounts for all
relevant symmetries in the $t$ channel, the GPs thus obtained must
represent the minimal set of GPs, even in the original $s$ channel.

As before, angular momentum conservation along the helicity axis
requires $|\lambda_1 - \lambda_2| \le J$.  It is convenient to
separate into the cases $\lambda_1 = \lambda_2$, $\lambda_1 =
-\lambda_2$, and all other pairs $(\lambda_1, \lambda_2)$.  If
$\lambda_1 = \lambda_2 \equiv \lambda$, then the $PC = ++$ and $PC =
-+$ cases in Eq.~(\ref{hel}) simplify to
\begin{equation} \label{cases}
\left[ \psi_{\lambda \lambda} + \psi_{-\lambda -\lambda} \right]
[ 1 + (-1)^J ] , \
\left[ \psi_{\lambda \lambda} - \psi_{-\lambda -\lambda} \right]
[ 1 + (-1)^J ] ,
\end{equation}
respectively, meaning that each is allowed only for $J$ even.  The
case $\lambda_1 = -\lambda_2 \equiv -\lambda$ is distinguished only
for $J^{-+}$, which according to Eq.~(\ref{hel}) vanishes identically.
All other $C=+$ helicity amplitudes are allowed for either parity as
long as $|\lambda_1 - \lambda_2| \le J$.

It then becomes a straightforward exercise in counting the values of
$\lambda_1, \lambda_2$ satisfying these conditions, and due to the
form of Eq.~(\ref{hel}) one may take $\lambda_1 \ge 0$ without loss of
generality.  Table~\ref{table1} summarizes the $t + \bar t$ helicity
constraint on each $J^{PC}$ and the number of allowed pairs
$(\lambda_1, \lambda_2$), which equals the number of allowed
amplitudes.  In cases where more than one value of $|\lambda_1 -
\lambda_2|$ is allowed, the number of amplitudes has been expanded to
show how many occur for each value.  All that remains to obtain a
final answer is to note that $[j] = j - [1-(-1)^{2j}]/4$ for $j$
integral or half-integral.  Summing the cases, one finds
\begin{equation} \label{GPnew}
\mbox{\rm Number of independent GPs} = 10s + 1 + \delta_{s,0} \ .
\end{equation}
Thus, $8s + \delta_{s,0}$ of the GPs in the counting leading to
Eq.~(\ref{GPold}) are redundant due to crossing symmetry.

\section{Counting RCS Helicity Amplitudes}

The final counting we perform is that of helicity amplitudes of RCS
for a target of spin $s$.  This again is carried out most conveniently
in the $t$ channel using $J^{PC}$ invariance (it can also be done in
the $s$ channel, where one must then impose time reversal invariance
since the initial and final on-shell photons are identical).  Helicity
amplitudes sum over many partial waves, meaning that the precise value
of $J$ is irrelevant to us as long as it is equal in the initial and
final states.  However, as is apparent from Eq.~(\ref{hel}), it gives
different linear combinations depending on whether $J$ is even or odd,
and these correspond to distinct helicity states.  This is apparent
from the discussion surrounding Eq.~(\ref{cases}): Since real photons
have only $\lambda = \pm 1$, there are two 2-photon helicity states
with $J^{++}$ even, $\lambda_1 = \lambda_2 = +1$ and $\lambda_1 =
-\lambda_2 = +1$, while only the latter is allowed for $J^{++}$ odd.
The $J^{-+}$ two-photon combination vanishes for $\lambda_1 =
-\lambda_2 = +1$, while for $\lambda_1 = \lambda_2 = +1$ it survives
only for even $J$.

To count the allowed $t + \bar t$ helicity combinations, we note that
it is sufficient to consider the two cases ($\lambda_1 > 0$ and
$\lambda_2 \ge 0$), which gives $[s+1/2] [s+2]$ combinations, or
($\lambda_1 = \lambda_2 = 0$), which gives $[1+(-1)^{2s}]/2$
combinations.  Again using the reasoning around Eq.~(\ref{cases}), if
$PC=++$, then the first case always allows $J$ even, and allows $J$
odd except when $\lambda_1 = \lambda_2$, which occurs in $[s+1/2] +
[1+(-1)^{2s}]/2$ combinations.  If $PC=-+$, then $J$ odd is not
allowed since the corresponding two-photon state vanishes, while $J$
even is allowed unless $\lambda_1 = -\lambda_2$, which occurs in
$[s+1/2] + [1+(-1)^{2s}]/2$ combinations.  Matching with the allowed
two-photon states, one finds a total of
\begin{eqnarray}
& & 4 \left\{ [s+1/2] [s+2] + [1+(-1)^{2s}]/2 \right\} - 2 \left\{
[s+1/2] + [1+(-1)^{2s}]/2 \right\} \nonumber \\
& = & 2(s+1)(2s+1)
\end{eqnarray}
RCS helicity amplitudes.  It is amusing to note that this number is
greater than the number of GPs in Eq.~(\ref{GPnew}) by an amount
$(2s-1)^2 - \delta_{s,0}$, a perfect square that vanishes for $s=0$
and 1/2 but not for $s \ge 1$.

In summary, we have shown that the number of independent generalized
polarizabilities for Compton scattering on a target of spin $s$ is
$(10s+1+\delta_{s,0})$, that the number of multipole amplitudes in the
soft photon limit, not taking into account the crossing symmetry of
the target, is $(18s+1+2\delta_{s,0})$, and that the number
$2(s+1)(2s+1)$ of real Compton scattering helicity amplitudes exceeds
the number of generalized polarizabilities for $s \ge 1$ and is equal
in the cases $s=0$ and 1/2 previously studied.

\acknowledgments
I would like to thank Richard Jacob and Xiangdong Ji for valuable
criticisms and illuminating conversations.  This work was supported in
part by the Department of Energy under grant No. DOE-AC05-84ER40150.

\begin{table}
\caption{Constraints on $t + \bar t$ helicities and number of
amplitudes for allowed values of $J^{PC}$.} \label{table1}

\begin{center}
\begin{tabular}{ccccc}
\hline\hline
$0^{++}$ && $\lambda_1 = \lambda_2$       && $[s+1]$   \\
$1^{++}$ && $|\lambda_1 - \lambda_2| = 1$ && $[s+1/2]$ \\
$2^{++}$ && $|\lambda_1 - \lambda_2| = 0, 1, 2$ && $[s+1] + [s+1/2] +
[s]$ \\
$0^{-+}$ && $\lambda_1 = \lambda_2 \neq 0$ && $[s+1/2]$ \\
$1^{-+}$ && $|\lambda_1 - \lambda_2| = 1$, \ $\lambda_1 \neq
-\lambda_2$ && $[s]$ \\
$2^{-+}$ && $|\lambda_1 - \lambda_2| = 0, 1, 2$ && $[s+1/2] + [s] +
\left( [s-1/2] + \delta_{s,0} \right)$
\\ \hline\hline
\end{tabular}
\end{center}

\end{table}

\end{document}